# Turbulent coherent structures and early life below the Kolmogorov scale

Madison S. Krieger [1]✉, Sam Sinai[1] & Martin A. Nowak [1,2]

Major evolutionary transitions, including the emergence of life, likely occurred in aqueous environments. While the role of water's chemistry in early life is well studied, the effects of water's ability to manipulate population structure are less clear. Population structure is known to be critical, as effective replicators must be insulated from parasites. Here, we propose that turbulent coherent structures, long-lasting flow patterns which trap particles, may serve many of the properties associated with compartments — collocalization, division, and merging — which are commonly thought to play a key role in the origins of life and other evolutionary transitions. We substantiate this idea by simulating multiple proposed metabolisms for early life in a simple model of a turbulent flow, and find that balancing the turnover times of biological particles and coherent structures can indeed enhance the likelihood of these metabolisms overcoming extinction either via parasitism or via a lack of metabolic support. Our results suggest that group selection models may be applicable with fewer physical and chemical constraints than previously thought, and apply much more widely in aqueous environments.

[1] Organismic and Evolutionary Biology, Harvard University, Cambridge, MA, USA. [2] Department of Mathematics, Harvard University, Cambridge, MA, USA. ✉email: mkrieger@fas.harvard.edu

At major transitions in the history of life, complexity has emerged via cooperation[1–4]. Most saliently, cooperation between groups of individuals is considered important for the origins of life, as well as the emergence of multicellularity[3–8]. One approach to formalize these events generally has been to define "coming together" as the aggregation of independent individuals, and "staying together" as the event where offspring remain close to their ancestors[4]. However, the emergence of cooperation at each of these transitions includes challenges that can be primarily ascribed to the exploitation of cooperators by parasites[6,7,9–11]—if cooperators interact and provide benefits to individuals in the population with equal probability, i.e. the population is well-mixed, those that receive benefits but do not reciprocate gain a competitive advantage and drive cooperators to extinction, thereby preventing the emergence of more complex entities[11]. To resolve this conflict, population structure models for early life were introduced in which individuals selectively interact with others[6,11–13]. These include lattice models, where individuals are restricted to interactions with particular neighbors (e.g. on a 2D surface)[13–16], and group (or multi-level) selection models[17], where individuals only interact with elements within their groups[6,11–13,18–20]. In particular, these models[6,10] and recent experimental studies[21,22] have been most intensively applied to problems in early biology and the origin of life. This is perhaps a natural domain of application because groups may be defined as concrete physical structures (e.g. droplets or protocells[5]) and the complexity of the underlying processes is relatively minimal.

In the context of abiogenesis, the study of spatial lattice models showed that cooperation can be maintained in spiral waves on simple 2D structures without flows[13,14]. One physical realization of these models can be envisioned as rocky surfaces potentially undergoing wet–dry cycles[13,15,23,24]. Group selection research, studied in abstract and also in the context of protocells, suggested that compartments provide necessary functionalities like collocalization of members for reactions[5,6,25,26], creation of gradients across boundaries (e.g. in lipid membranes[27]), exclusion of parasites through division[11,19,20] and rise of diversity through merging[28], all provide benefits for cooperators.

Many populations of biological organisms exist in an aqueous environment. As such, additional physical mechanisms governing population structure which are compatible with aqueous environments, such as active droplets[5], slicks of fatty oils[29], bubbles containing aerosols particles[30], or surfactant micelles[31] have been considered as early mechanisms for group selection. Interestingly, while the chemical role and necessity of water itself for emergence and maintenance of life is well-appreciated, the potential role of its transport properties (in the absence of additional structures such as those described above) is relatively unexplored. Recent years have seen the beginning of interest in how flows affect population genetics[32–38], but few works address cooperation, which depends much more fundamentally on population structure than does the spread of (dis)advantageous alleles[39,40]. As has been described recently[34,35], the well-mixed assumption in ecology (an assumption that a population may be represented by a single nonspatial compartment, i.e., all individuals interact with all other individuals at all times) may never be realized in fluid flows, even those which are strongly mixing or ergodic. It is therefore sensible to ask whether the types of population structure which appear naturally in fluids are conducive to the types of cooperation necessary for early evolutionary milestones.

To address this question, we consider the functionalities that motivated classical group selection and ask if moving fluids can reproduce them. We explore in particular if (I) collocalization, (II) division, and (III) merging, which are physical properties of groups, can be replicated. Because many of the aqueous environments pertinent to life are not quiescent, such as the surface of the ocean, ponds, or montane streams, we test the hypothesis that turbulent flows can support these fundamental aspects of early group selection.

Our interest in these questions is therefore closely related to the recent surge of interest in the stirring and mixing of passive tracers in turbulence, and especially related to the dynamics of coherent Lagrangian structures[41–44] (Lagrangian coherent structures (LCSs)), long-lasting material surfaces within a flow which roughly divide up the spatial domain into regions with very different transport properties.

The Lyapunov exponent (which measures the convergence or divergence of nearby fluid particle trajectories), measures the "skeleton" of the flow, identifying surfaces that are highly repelling, attracting, or elliptical (meaning particle trajectories remain parallel for long times)[45,46].

LCSs can be thought as isolated regions which trap fluid and thus acts like compartments necessary for group selection. Furthermore, to the naked eye they seem to replicate the qualitative features (I–III) above which are necessary for evolution via cooperation—in real flows, one can observe the creation and destruction, merging, and division of LCSs. They tend to arise spontaneously whenever the fluid is put in motion by some large-scale forcing, like surface winds, and they are eventually destroyed by viscous dissipation[47,48].

In the rest of this paper, we substantiate the idea that LCSs in fluids can provide the features (I–III) above (collocalization, division, and merging) that would have been impactful in early cooperative stages of evolution. We study these structures by performing simulations of replicating cooperators embedded within turbulent flows. Furthermore, once these are established, we demonstrate that the role of these group-like properties also contribute to the spread of genetic diversity.

## Results

**Modeling emerging populations in turbulent flows.** We consider a population of biological organisms in a moving fluid (Fig. 1). We are particularly interested in the problem of cooperation at very small scales, from tens of nanometers (e.g., small replicating RNA strands) to microns (e.g., single-celled organisms). Even at these small distances, the fluid can be described as a continuum, because the mean free path of water molecules (the average distance traveled without a collision with another molecule) is roughly an angstrom. Thus the Knudsen number (the ratio of the mean free path to the size of the particle) for the smallest object of interest would be no larger than $10^{-2}$. This means that even the smallest biological particles under consideration undergo many collisions with the surrounding fluid molecules before they can travel an appreciable distance.

We furthermore suppose that the fluid under consideration undergoes some turbulent motion, as this can be expected to occur at least intermittently in most aqueous environments. While turbulent flows are perhaps best known for their ability to enhance mixing via erratic motion[49], their habit of doing the opposite, creating coherent spatial structures, is sometimes overlooked. However, at nearly all spatial scales, regions of both enhanced and inhibited mixing can be found in a turbulent flow.

This duality of turbulent behavior is perhaps best measured by the finite-time Lyapunov exponent (FTLE). The FTLE, $\lambda$, estimates the rate at which two particles starting a distance $\delta r$(0) drift apart

$$\lambda = \log\left(\frac{\delta r(t)}{\delta r(0)}\right), \qquad (1)$$

where $\delta r(t)$ is the particle pair separation at some time $t$ after $t = 0$. Regions of positive FTLE characterize the chaotic behavior of turbulence which leads to enhanced dispersion. Regions of zero and negative FTLE, on the other hand, are regions where passive tracers (inert, massless particles that simply follow the flow) are approaching or staying near one another for a time interval $t$. When such regions persist in time they form the LCSs discussed in the "Introduction" section. LCSs with scales of hundreds of kilometers can be seen in satellite pictures of ocean color, because they trap colonies of chlorophyll-rich plankton[50]. But they also exist at smaller and smaller scales, down to the Kolmogorov lengthscale $\eta$, below which viscous forces eventually become so strong that they dissipate any coherent structure. For reasonable aqueous environments, the Kolmogorov scale is on the order of tens or hundreds of microns up to millimeters, much larger than our biological particles of interest. When biological particles are in regions of positive FTLE, they mix chaotically; when they are in regions of negative FTLE, they remain close to their neighbors for long times due to advective forces.

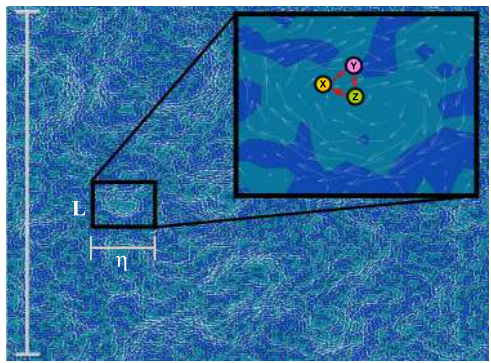

**Fig. 1 Compartmentalization appears naturally in fluid flows.** Segregation of particles into different regions can occur in two-dimensional flows for a variety of reasons. Here, we posit that coherent structures, which are a common spatiotemporal motif in turbulence, could provide long-lived safe havens for cooperators. Here we show an example of a two-dimensional fluid flow on a periodic domain, where coherent vortices have formed from random initial conditions (white arrows: velocity vectors, blue shade: local vorticity). The size of the whole domain is $L$, and the size of the smallest coherent spatial structures is the Kolmogorov lengthscale, $\eta$. White arrows show the velocity field, and colored shading shows vorticity contours. The inset indicates the local flow field near a large vortex, along with individuals from three different species ($X$, $Y$, $Z$) in a cooperative metabolism. Here we show a simple cooperative hypercycle, in which $X$ catalyzes $Y$, $Y$ catalyzes $Z$, and $Z$ catalyzes $X$. While all particles die at the same rate $d$, birth rates increase under catalyzation, which occurs when a catalyzing particle is within a distance less than the interaction radius $R_{int}$ of another particle.

We will also neglect the role of thermal diffusion inside the LCSs. We ground this in a comparison between the roles of advection and thermal diffusion at the lengthscales of interest. The relevant lengthscale for diffusion is the Batchelor scale, $\lambda_B$, which gives the approximate distance a particle with diffusion constant $D$ will drift due to Brownian motion as it is advected a distance on the order of the Kolmogorov scale[51]. The ratio of the Batchelor (diffusive) lengthscale to the Kolmogorov (advective) lengthscale is given by the Schmidt number, $\lambda_B/\eta = Sc^{-1/2} = \sqrt{D/\nu}$, where $\nu$ is the kinematic viscosity of the fluid. In water, $\nu = 10^6 \, \mu m^2 \, s^{-1}$. An upper bound for $D$ for the biological objects of interest can be given by the diffusion rate of a single nucleotide in water, since these early lifeforms would be at least as large as many linked nucleotides and would therefore have smaller diffusion rates. For a single nucleotide, a measurement of $D \approx 400 \, \mu m^2 \, s^{-1}$ has been given[52]. Therefore, the upper bound on the Batchelor scale is roughly 2% the size of the Kolmogorov scale, meaning diffusion represents only a very small correction to the advective displacement. As a first approximation, then, we neglect the role of Brownian motion, and we revisit the possible errors incurred in Supplementary Note 2 and Supplementary Fig. 3. Note that at smaller scales, when chemical compounds were forming nucleotides, diffusion constants would be higher and therefore diffusion more important—both thermal diffusion and advective mixing in a model similar to ours have shown greatly enhanced reaction rates in chemical systems[53].

For the global fluid flow, we use a minimal two-dimensional point-vortex model[54], that has been previously employed as a toy model for homogeneous turbulence[54–56]. We consider a set of $N$ vortices of strength (circulation) $\Gamma_j$ and position $\mathbf{z}_j = (x_j(t), y_j(t))$. The fluid flow $\mathbf{u}(x,y) = (u_1(x,y), u_2(x,y))$ at a point $\mathbf{x} = (x,y)$ is given by

$$u_1(\mathbf{x}) = -\frac{1}{2\pi} \sum_{j=1}^{N} \frac{\Gamma_j (y - y_j)}{\|\mathbf{x} - \mathbf{x}_j\|^2}, \qquad (2)$$

$$u_2(\mathbf{x}) = \frac{1}{2\pi} \sum_{j=1}^{N} \frac{\Gamma_j (x - x_j)}{\|\mathbf{x} - \mathbf{x}_j\|^2}. \qquad (3)$$

Because the point vortices are massless, conserved features of the flow, their positions are also advected by the flow (and therefore their positions $\mathbf{z}_i(t)$ are given by Eqs. (2) and (3), with the left-hand side terms replaced by $\frac{d}{dt} x_i(t)$ and $\frac{d}{dt} y_i(t)$, respectively). For simplicity, we assume uniform strength among vortices $\Gamma$, but consider clockwise ($\Gamma < 0$) and counter-clockwise ($\Gamma > 0$) vortices. The vortices are confined to meander within a square box of size $L$. To avoid the issues of edge effects, we assume that our domain is doubly periodic (see "Methods" section). The simplicity of our model aids in capturing the essential qualitative and statistical features of turbulence, and does not depend on the particular mechanisms that result in conserved vortices.

One major advantage of this point-vortex model is that, in addition to being computationally efficient to solve (see "Methods" section), it reproduces the coherent structures discussed in the "Introduction" section (trapping regions with finite lifespans, arising between the point vortices—the vortices are not themselves coherent structures). We could just as well have considered a fully three-dimensional flow, as the coherent structures we are interested in also appear in three spatial dimensions[57,58]. However, this would require not only a more complicated fluid model, but would also reduce the generality of our results. For instance, considering a three-dimensional flow forces us to decide to what extent the fluid should be stratified, which would restrict us to a (an)haline environment. Considering partial effects of a third spatial dimension (such as apparent compressibility and the associated upwellings and downwellings, which have been considered in other works[36]) also force us to make additional assumptions about the biology, since we must decide if a particle caught in a downwelling is destroyed or can re-emerge in an upwelling, which may implicitly make assumptions about the environmental and biochemical requirements of these organisms. While these additional kinematics are no doubt relevant, we have left these complications to future work which may focus on different fluid environments (the ocean, warm little ponds, fast streams, etc.) and their potential (dis)advantages in the context of our results derived here on the potential importance of coherent structures.

The point-vortex model is also appealing because it qualitatively captures features of turbulence, having spatiotemporal regions of both fast and slow flows, with material surfaces that are attracting, repelling, and neutral[54,59]. In actual applications, the

| Table 1 Reactions for the various metabolisms used in this work. | |
|---|---|
| **Metabolism** | **Birth reactions** |
| Replicase R1 | $A \xrightarrow{s+\beta} 2A$, $(A+A) \xrightarrow{s+\beta} 3A$, $B \xrightarrow{s} 2B$, $(A+B) \xrightarrow{s+\beta} A+2B$ |
| Replicase R2 | $A \xrightarrow{s} 2A$, $(A+A) \xrightarrow{s+\beta} 3A$, $B \xrightarrow{s} 2B$, $(A+B) \xrightarrow{s+\beta} A+2B$ |
| Hypercycle | $A \xrightarrow{s} 2A$, $(A+B) \xrightarrow{s+\beta} 2A+B$, $B \xrightarrow{s} 2B$, $(A+B) \xrightarrow{s+\beta} A+2B$ |
| While all particles die at the same rate $d$, birth rates ($s$ in absence of catalyzation) increase under catalyzation ($s \to s+\beta$), which occurs when a catalyzing particle is within a distance ($\ell$) less than the interaction radius $R_{int}$ of another particle it has the ability to catalyze. In the replicase models, particle $A$ can catalyze both $A$ and $B$, but $B$ cannot catalyze $A$—the difference between R1 and R2 is whether a single $A$ particle can (R1) or cannot (R2) catalyze itself. In the two-member hypercycle shown here, $A$ can catalyze $B$ and $B$ can catalyze $A$, but $A$ particles cannot catalyze other $A$ particles, nor can $B$ particles catalyze other $B$ particles. In the $n$-member hypercycle, $A$ particles can only catalyze $B$ particles, $B$ particles can only catalyze $C$ particles, and so on, with the $n$th member only being able to catalyze $A$ particles. | |

strength $\Gamma$ and the density of vortices (set by $N$, keeping the domain size fixed) would be tuned to account for the specific properties of the flow under consideration, but the qualitative results described below are independent of these specific choices.

Note that the point vortices which generate the flow are permanent; only the coherent structures which arise around and between them are impermanent. The permanence of the point vortices allows the flow to propagate forever without a change in its statistical properties, which is desirable since the time-independence of the model statistics is necessary to compare the impact of flow versus biology timescales in different simulations using different biological parameters.

The biology of the particles is modeled by a stochastic birth–death process. Besides advection by the flow, particles can undergo one of two reactions: death or replication. Replication is defined as the appearance of a new particle, of the same species as the parent particle (described below), in the vicinity of the parent. Both birth and death are modeled by Poisson processes: a given particle dies between time $t$ and time $t + dt$ with probability $d$, so that in a very large population, the total size is decreasing with a rate of roughly $d$ (we set the death rate as a constant). The replication rate, however, is implicitly time-dependent and space-dependent, because of the possibility of cooperation.

We consider different models of early cooperative metabolisms from previous studies[10,60]. These models are summarized in Table 1. To incorporate spatialization, we consider an effective radius of interaction, $R_{int}$, within which a individual can provide metabolic benefits to another. Our metabolisms under consideration capture most of the critical dynamics in early cooperation that might arise, no matter the exact chemical or physical pathway. We follow the naming convention for different metabolism situations[10]: in "Replicase R1", the replicase species $A$ cooperates with any other particle, but also itself, so its replication rate is effectively space-independent (we assume the benefits of cooperation are not additive, so that a particle at any time is either in a state of being cooperated—with or not). All other species ($B$, $C$, etc.) are effectively parasites. "Replicase R2" differs from "Replicase R1" only insofar as a replicase particle $A$ cannot cooperate with (replicate) itself (though distinct particles of the replicase type can still cooperate with one another). Finally, we study a hypercycle, representing a system in which an individual of type $A$ can only cooperate with $B$, $B$ with $C$, $C$ with $D$, and so on, with the last type able to cooperate with $A$, which is a well-studied mechanism for an early metabolism[60,61].

We use a finite-population, agent-based approach where an emerging population would have faced very basic challenges to its survival; should these challenges be surmounted and the population grow to a much larger size, then issues we omit from our modeling (such as resource limitation, or the possibility of modeling the population as a continuous field) would become relevant. Some of these situations have already received attention in the literature[32,33,36–38]. Our work therefore bridges the gap between these works, which focus on constant-fitness populations at high numbers in turbulent fluids and the case where there are no organisms, by considering both small-population effects and frequency-dependent fitness. Both the hydrodynamic and biological aspects of our model are summarized in Fig. 2.

**Natural fluid timescales may impose selective pressures on early replicators**. To summarize the interplay between turbulent transport and biology we introduce the Damköhler number $Da$, which is the ratio between the timescale over which the velocity experienced by a particle changes and the characteristic timescale of the biological process under consideration. The Damköhler number has proven useful in other studies of biological processes in turbulence[62], where global extinction can be guaranteed beyond a certain threshold value.

The typical free path of a particle in a flow field generated by point vortices is given by the mean inter-vortex distance, because particles will change direction every time they collide with a new vortex. The inter-vortex distance in our setup is given by $\xi \propto L/\sqrt{N}$. The characteristic velocity experienced by particles in this flow is given by $\Gamma/2\pi\xi$. A characteristic timescale for the trajectories is therefore given by distance over velocity $\tau_F \propto L^2/N|\Gamma|$. This is essentially the time for which a particle travels in the fluid before colliding with a new vortex. A natural timescale for biology is the expected lifetime of a particle, which is simply $\tau_B = 1/d$. The Damköhler number is therefore $D_a = \tau_F/\tau_B = dL^2/N|\Gamma|$. In our simulations we scale time on $1/d$ and length on $L$, so that effectively $Da = (N|\Gamma|)^{-1}$.

Examining the interplay of physics and biology at extreme Damköhler numbers reveals how biological and physical time-scales must be in relative harmony for cooperation to flourish in the absence of additional mechanisms, such as a lipid membrane. At very high Damköhler numbers, particles undergo millions of generations before being dispersed by the fluid. Particles initialized in a coherent structure will remain there, along with any parasites initialized nearby; should they survive, their lineage will struggle to spread even by migratory events, since the transit time between coherent structures becomes so long that a large number of comigrating cooperators is necessary to ensure survival. As seen in Supplementary Movie 1, an initially randomly distributed population (here, a two-species hypercycle) simply coarsens into a few dense clusters, with others dying out before they can be brought within distance of other cooperators. High Damköhler number environments therefore satisfy the requirement of "staying together"—however, without the ability to divide and merge over biologically significant timescales, avoiding parasites (which we may imagine emerge via mutations and therefore arrive at a fairly regular rate in any population) would be quite challenging.

Extremely low Damköhler numbers (Supplementary Movie 2) are also detrimental to survival. In this limit, particles disperse quickly throughout the flow on biological timescales, undermining cooperation. We present relevant statistics for this limit in Supplementary Fig. 2. Most particles spend the majority of their expected lifetime (in the absence of catalytic aid) alone, and do not benefit from cooperation. Indeed, in simulations with small inocula and low Damköhler numbers we always observe extinction.

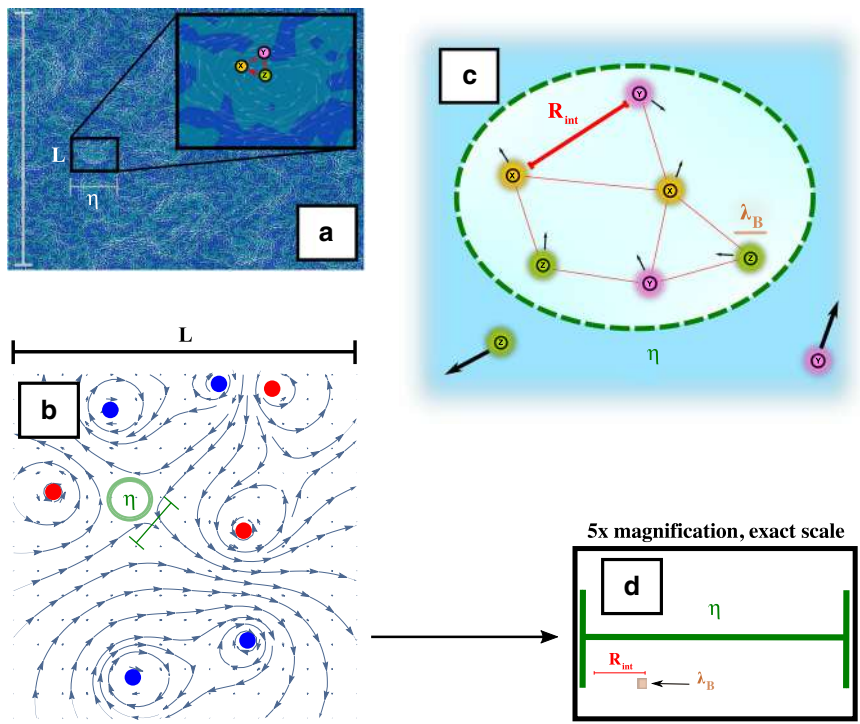

**Fig. 2 A summary of the model. a** We consider the situation of cooperative catalyzation in a turbulent fluid. For computational tractability we restrict our attention to two dimensions, since the trapping regions (Lagrangian coherent structures, LCS) which are of qualitative interest arise here from simple models. We consider the full domain of interest to be of size $L$, with $L \gg \eta$, where $\eta$ is the Kolmogorov scale, the scale of the smallest advective spatial structures in turbulence. **b** Our modeling of the flow. Many qualitative as well as quantitative features of turbulence are well-captured by a toy model consisting of multiple point vortices (here, 7 vortices: blue = clockwise, red = counterclockwise). To avoid edge effects, the domain is doubly periodic. In this snapshot, there is a quiescent region circled in green, whose size represents the Kolmogorov scale. If this spatial region remains quiescent for long times, effectively trapping particles (exhibiting a negative finite-time-Lyapunov-exponent), it will represent an LCS. **c** Within an LCS, particles separate much more slowly from one another than outside (black arrows). Over the lifetime of this LCS, the particles will, in addition to advection, diffuse a distance given by the Batchelor scale $\lambda_B$, and so we could consider them "blurred" over this distance. For a nucleotide in water, $\lambda_B \approx 0.02 \times \eta$. Biological particles (colored X,Y,Z) can catalyze each other if they are within a distance $R_{int}$, and therefore a single snapshot induces a dynamic graph structure of cooperators and defectors. **d** ×5 magnification of $\eta$ as in **b** as well as the other relevant lengthscales drawn to scale; here we show the value $R_{int} = 0.03*L$ typically used in our simulations (see "Methods" section), and $\lambda_B = 0.02\eta$ (thickened to a box to aid the eye).

To interpret results on a whole range of Damköhler numbers, it is useful to introduce the two-particle covariance $G(x_1, x_2)$, defined as the probability that a pair of particles is found at positions $x_1$ and $x_2$. Averaged over many realizations, our process is isotropic and homogeneous, and the pair covariance is only a function of the interparticle distance rather than the exact positions, so that the probability of a pair separated by a distance $\|\mathbf{x}_1 - \mathbf{x}_2\|$ is $G(\|\mathbf{x}_1 - \mathbf{x}_2\|)$. We plot a relevant example of this quantity in Fig. 3 for three cases where: the transport and biological timescales are comparable ($Da = \mathcal{O}(1)$); where there is no fluid motion and $\Gamma = 0$ ($Da \gg 1$); and where there is no biological effect and $d = 0$ ($Da = 0$). The relative contributions from biology and flow have competing effects, each with clear advantages and disadvantages for cooperation. In the absence of motion, reproducing populations will always have the largest values of $G(\|\mathbf{x}_1 - \mathbf{x}_2\|)$ at the smallest values of $\|\mathbf{x}_1 - \mathbf{x}_2\|$, since birth places offspring nearby, whereas death can occur anywhere, leading to cluster formation[63,64]. This means that a large number of pairs are within the interaction radius $R_{int}$—a great benefit in the absence of parasites, but a serious liability in the presence of even a few parasites. On the other hand, with increasing motion, pairs are found at increasing distances, which means that with a small value of $R_{int}$, very few particles are interacting and the population is expected to go extinct.

Intermediate Damköhler number offer a favorable trade-off, because motions prevent particles from aggregating completely at small separations (which protects against parasites, as we show below) without completely eliminating the possibility for interactions. Interestingly, the importance of intermediate Damköhler number for population fitness has already been observed in experiments involving bacterial mutualism where the bacteria are also motile[65].

**Coherent structures collocalize replicating particles.** So far we have focused on a rough estimate of the rate at which two particles drift apart from each other, $\tau_F \propto L^2/N|\Gamma|$. However, in a turbulent flow, like the one generated by the ensemble of point vortices, particles pairs are occasionally trapped into LCSs and remain close to each other for much longer than $\tau_F$ until the LCS breaks apart and dispersion resumes.

As a quick illustration that collocalization naturally occurs in LCSs, we zoom in on one and model the flow within as a purely circular flow, subject to an oscillatory perturbation of strength $\varepsilon$, a flow which is often studied in work on coherent structures[66–69]. In the absence of this perturbation, the FTLE will be zero, as we expect from an LCS. However, as the perturbation strength

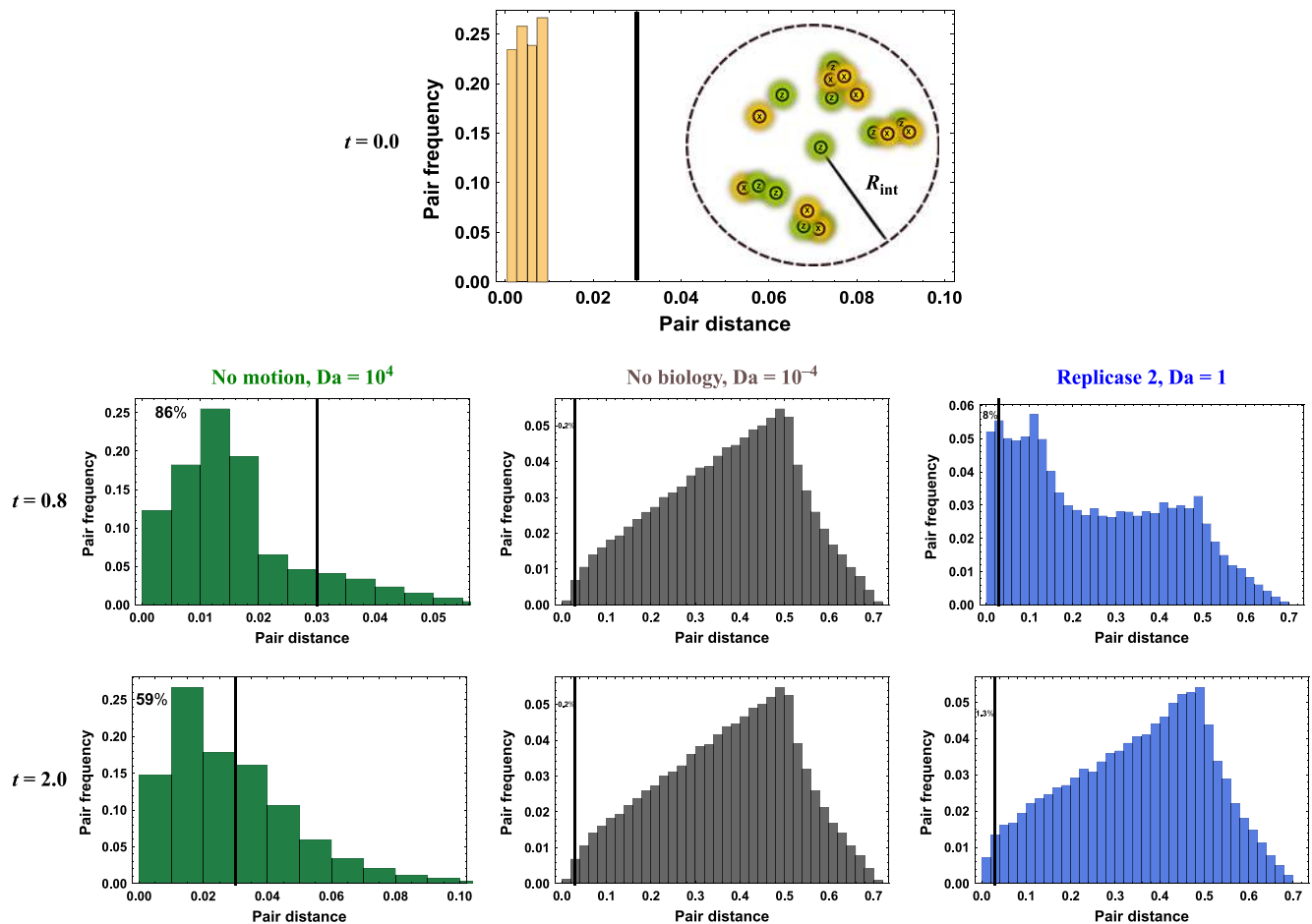

**Fig. 3 The relative value of the fluid velocity and metabolic timescales is critical for cooperation.** The importance of the Damköhler number, and of flows in general, can be understood via the pair covariance $G(|x_1 - x_2|)$, which gives the probability of finding a pair with interparticle separation $|x_1 - x_2|$. Starting from an initial condition of many particles in close proximity (so $G(|x_1 - x_2|)$ can be approximated by a delta function at $G(0)$), the evolution of pair covariance is governed by competition between flow and biology. Histograms show an average over 1000 simulations. Times are measured in the expected lifetime of a single biological particle. When particles reproduce, we use the interaction radius $R_{int} = 0.03$, and the fraction of interacting particles (left of black bar) is given at the top-left. In the absence of flow (green histograms), the initial condition will slowly spread into a large colony, with most particles within the interaction radius, leaving the susceptible to parasites. On the other hand, passive tracers in a turbulent flow (gray histograms) obey a known power law[84], in which the interparticle distance increases on average, here quickly reaching a limiting distribution due to the doubly periodic nature of our spatial domain. In this situation, so few pairs are within the interaction radius that most lineages should be expected to die out and the population on average goes extinct. A replicating population in a flow at $Da = \mathcal{O}(1)$ (blue histogram), however, can combine the advantages of both situations, creating rich structure ($G(x_1, x_2)$ having broad support) while also having a higher number of interacting pairs than in the case of no biology.

increases, the flow becomes more chaotic, eventually dissolving the coherent structure.

To illustrate collocalization, we considered our representative metabolisms (Table 1) in this flow, as shown in Fig. 4. We asked how often a small inoculum of a metabolism, consisting of only a 10 particles, could successfully multiply to one thousand particles (establish), as the interaction distance $R_{int}$ and the chaotic perturbation strength $\varepsilon$ were varied. Increasing the value of $R_{int}$ effectively decreases the importance of spatial structure, as $R_{int} = 1$ represents a well-mixed population, in which a population made only of cooperators will always thrive. For lower values of $R_{int}$, the effect of the collocalization is important, with a purer coherent structure having clear advantage over a chaotic region. For instance, with $R_{int} = 0.01$, an inoculum of Replicase 2 would likely perish in a chaotic flow with no coherent structure ($\varepsilon = 1$), but inside a coherent structure ($\varepsilon = 0.01$) it multiplied from ten particles to a thousand in roughly 40% of simulations.

**Segregation of coherent structures can insulate cooperators from parasites.** While collocalization is necessary for cooperation to take place, it is prone to parasitism. If some elements become defective over time and cease providing cooperative support (effectively becoming parasites), the entire population of cooperators becomes vulnerable to collapse[9]. In compartment models, group selection has been offered as a solution to this problem, as selection works against groups that contain a large number of parasites.

Since turbulent flows naturally generate multiple LCSs, it would seem that group selection can also preserve cooperators by the same mechanism. However, the analogy is not perfect. LCSs are impermanent, even if they can be long-lived. Further, since they are not physical membranes, and merge and divide with some regularity, they allow a degree of mixing which is sometimes forbidden in theoretical compartment models. It therefore remains to be shown that LCSs provide the kind of population segregation necessary for cooperators to escape parasites.

To demonstrate the role of turbulent flow in isolating and removing defective elements, we consider the population dynamics of an R2-replicase in a system of $N = 7$ point vortices. All particles within range of a cooperator are catalyzed, even if there is a single cooperator. Furthermore, we do not enforce a resource limitation or maximum population size.

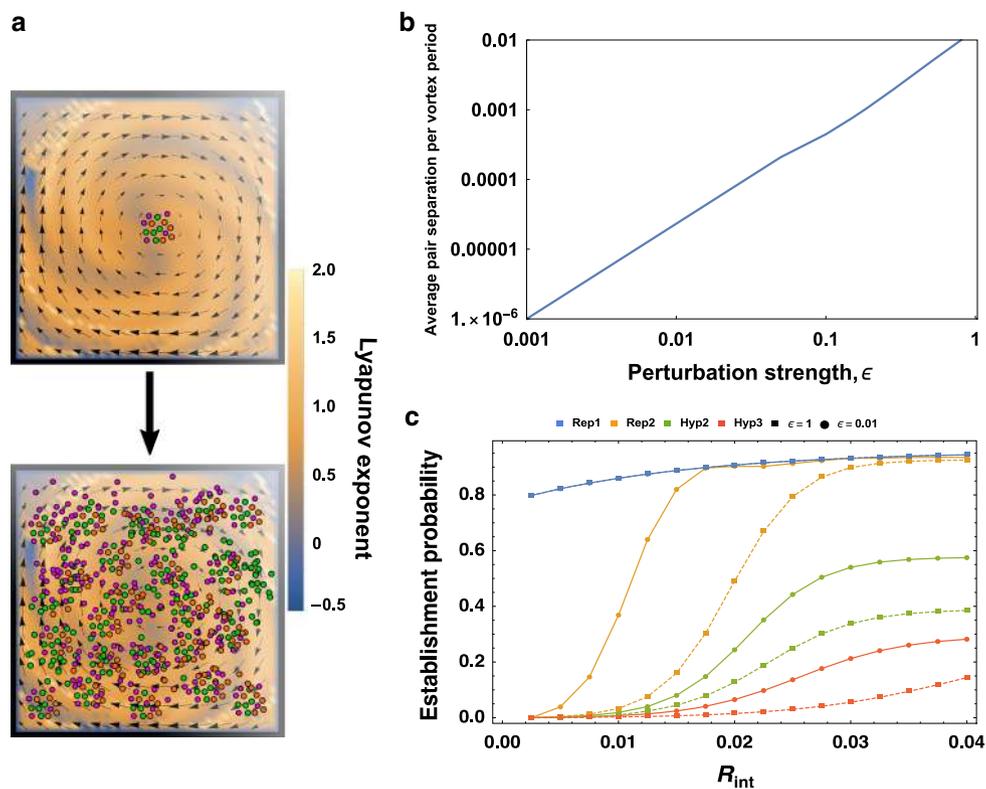

**Fig. 4 Coherent structures collocalize replicating particles. a** A sample velocity field consisting of a circular flow perturbed by an unsteady oscillation of amplitude $\varepsilon = 0.1$ atop a plot of the finite-time Lyapunov exponent, a common measure of the local "chaoticity" in a flow. As $\varepsilon$ increases, particles (colored dots, schematizing different species) become more likely to cross onto the unstable manifold and separate from nearby neighbors. We define "establishment" as a hundredfold increase in the population size starting from a very small initial condition, where continuum limits previously described[32,33,36] begin to be applicable. **b** The average particle separation (per flow period) of an initially adjacent pair of particles, versus the strength $\varepsilon$ of the perturbation, in logarithmic scale. **c** Results of numerical simulations of various metabolisms (colors) for varying distances $R_{int}$ at which particles can cooperate with one another. Square markers indicate the likelihood that the population increases 1000-fold starting from an inoculum of five particles of each species injected into a chaotic field (no coherent structure, $\varepsilon = 1$), whereas circular markers indicate the same likelihood for an identical inoculum injected into a nearly pure coherent structure ($\varepsilon = 0.01$). Note that for Replicase R1 (blue), the establishment probability is equal for both $\varepsilon = 1$ and $\varepsilon = 0.01$, so that only the solid line is shown.

In the well-mixed (i.e., no spatial structure; all individuals interact with all other individuals) limit of this system, cooperators and parasites are equally fit, and if the population size were to be fixed, with approximately equal fractions of both species, one species would go extinct with 50% probability simply due to stochastic effects. We first show the benefit of segregation then by considering a constant-population-size Wright–Fisher process[70], and simply count the number of realizations in which parasites go extinct before cooperators. If this number of realizations is higher in our spatial process than in the well-mixed process, then segregation due to flow is helpful to keep cooperators safe from parasites. Note that in our actual system (where population size is unrestricted), since we do not allow for mutations between species, parasites going extinct means the system flourishes; cooperators going extinct means the system collapses.

The results of these Wright–Fisher simulations are shown in Fig. 5. We found that cooperators can do slightly better than parasites for certain values of $R_{int}$ and $Da$, indicating a positive effect from the flow. As seen in previous work[34], the well-mixed results are seen at much lower values of $R_{int}$ than those which span the system, with the results already visually indistinguishable from the well-mixed prediction for $R_{int} \approx 0.1$.

We then lifted the restriction of fixed population size, simply tracking the relative sizes of the parasite and cooperator populations in our normal branching Replicase-R2 system. Over 1000 simulations conducted until the population had grown from size 50 to size 5000 (or extinction), we never once saw extinction. Rather, the fraction of cooperators seems to be roughly normal or possibly log-normal, suggesting that perhaps some fraction of realizations would eventually lead to extinction. This shows that the results from the fixed population size simulations only provide partial intuition for how cooperators and parasites become segregated. The reason why this intuition is only partial is because when parasites fix, the population goes extinct, rather than continuing on as a monomorphic parasite population. Additionally, in a population with dynamic size, one of the two absorbing states of the fixed-population model (either all-cooperators or all-defectors) need not be reached. The system can sustain a fraction of both cooperators and parasites, similar to what has been seen in other models of cooperation with fluctuating population size[71].

We also tracked the pair correlations between cooperator–cooperator and cooperator–parasite in these simulations. In aggregate (e.g., all possible pairs in all realizations, binned together), the cooperator–cooperator and cooperator–parasite pair correlation functions exactly match that which was shown in Fig. 4 (the maximum difference between any bin and the corresponding bin in Fig. 4 was <$10^4$). However, on a realization-by-realization basis, there seemed to be some correlation between the emerging of structures which segregated cooperators from parasites and their eventual fractions.

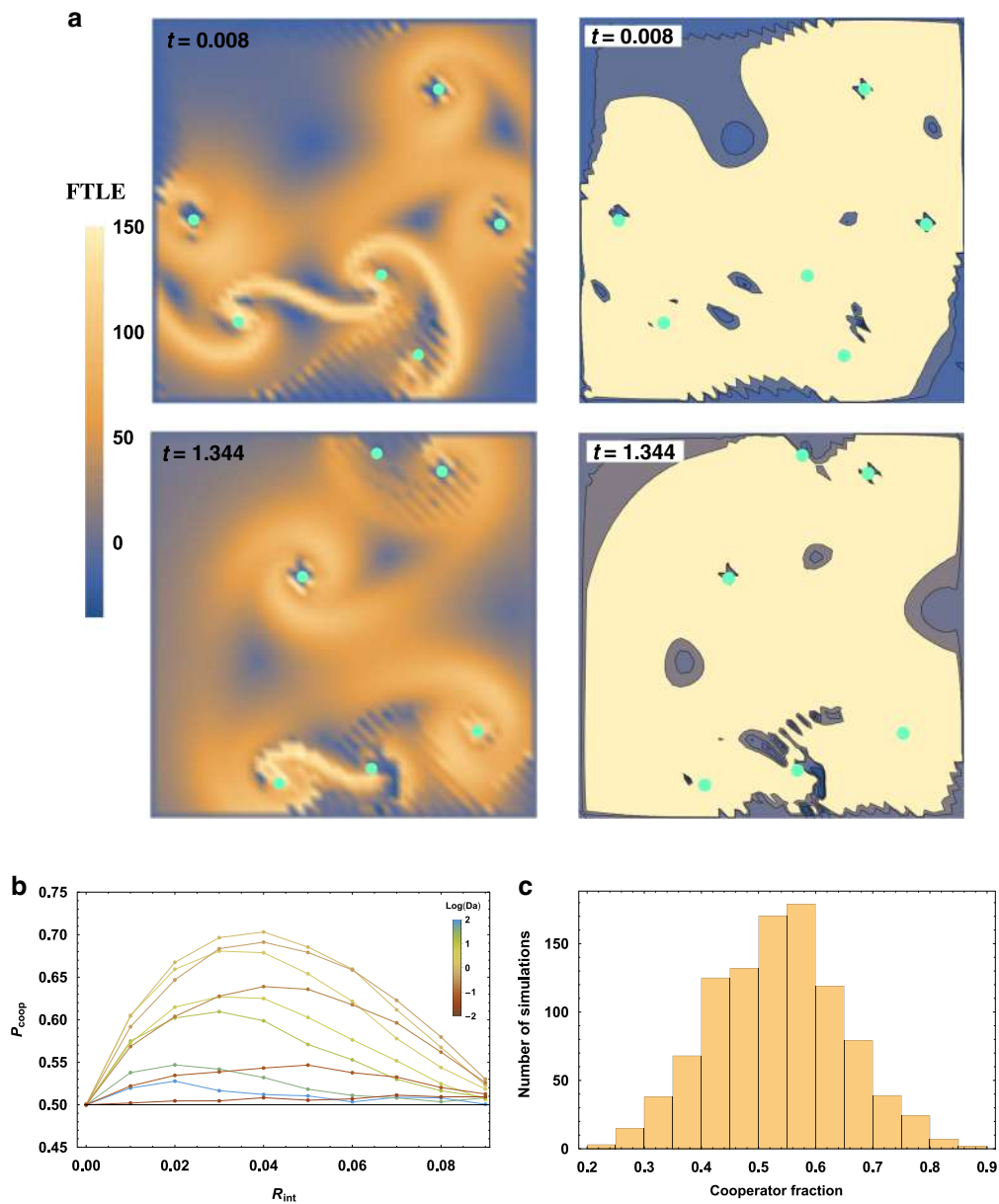

**Fig. 5 Trapping and segregation by coherent structures gives cooperators a boost.** In a two-species Replicase R2, one species (the replicase) plays the role of a cooperator (at no cost, provides a catalytic benefit of size $\beta$ to all non-self organisms within a radius $R_{int}$), while the other species plays the role of a parasite. **a** Chaotic flows generate LCSs with finite lifetimes, which act as traps for particles. Here a six-vortex (teal dots) example is shown with the finite-time Lyapunov exponent (FTLE) at left and trapping regions (negative FTLE) at right. **b** In a well-mixed population obeying a Wright–Fisher update rule on constant population size, each species would be equally likely to fix (reflected by $P_{coop}$, the fixation probability of cooperators shown on the y-axis, equaling 1/2) if they began at equal fractions. To test whether flow-segregation helped replicases, we performed Wright–Fisher simulations on a population of size $N = 200$ at different $R_{int}$ and Da. Intermediate values of Da show a boost for cooperation at specific values of $R_{int}$. **c** Returning to a birth–death process, where the population either goes extinct or grows to infinite size, we tracked the average replicase fraction of the population over 1000 simulations from $t = 0$ to $t = 10$, with roughly 50 members of each species at $t = 0$. We never saw either population go extinct when starting from even such a small size; rather, the population tends to sustain both parasites and replicases, with on average slightly more replicases than parasites.

**Flows create rich population structure via division and merging of coherent structures.** In the previous section, the existence of multiple LCSs facilitated compartmentalization of the population such that the deleterious effect of parasites was avoided. However, in order for LCSs to have been truly useful in early life, they must have further properties—namely, they must be able to permute their biological contents by division and merging.

As discussed in the "Introduction" section, LCSs continuously appear, merge and disappear in a turbulent flow as a result of nonlinear dynamics. Thus LCSs seem to replicate the qualitative features of division and merging even in the absence of life.

To investigate whether this might have induced useful division and merging on life, we compared the behavior of our processes and the behavior of passive tracers in the same flow. Comparing passive tracers and reproducing particles allows us to understand what properties of the demographics seen in the simulations are due purely to the flow and what properties are due to the biology, and also allows us to compare the relative importance of the two.

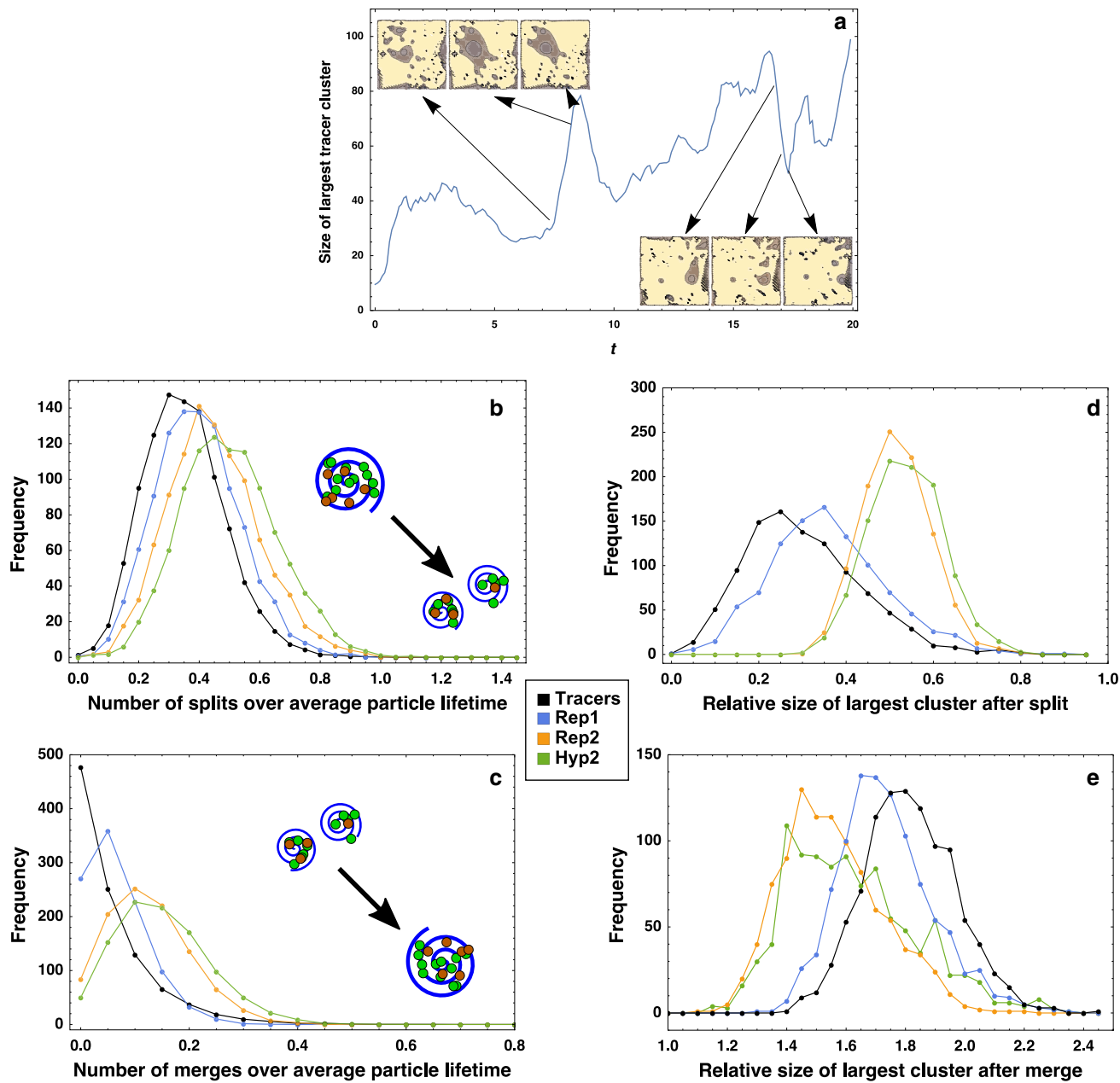

**Fig. 6 Merging and splitting of LCSs leads to rich, dynamic population structure.** Based purely on the birth–death aspect of our process, large changes in the size of the largest particle cluster are improbable; for instance, the probability of monotonic changes >10 or so particles is always <0.001%. In **a** this is highlighted by removing birth and death, leaving only passive tracers; large changes in cluster demographics can therefore only arise from the flow. Such changes are therefore likely due to dramatic events in the fluid Lyapunov exponent landscape, such as division and merging of LCSs. Restricting our attention to only these large demographic swings over one thousand realizations yields histograms on the number of **b** splits, **c** merges, as well as **d** the percentage decrease in size of the largest cluster after a split and the **e** percentage increase in size of the largest cluster after a merge. Dots show the center of histogram bins, while the value on the y-axis shows the height of the bin; full bars have not been drawn to aid visualization. While all statistics for different metabolisms can be roughly approximated by the behavior of passive tracers, the inherent patchiness arising from increasingly baroque metabolisms leads to larger merge and split frequencies with smaller average effects on demographics.

Analyzing passive tracer behavior also suggests when the landscape of LCSs is changing without necessitating the costly calculations to actually map the finite-time-Lyapunov exponent (indeed, many algorithms for detecting LCSs rely on analyzing the dynamic-graph properties of passive tracers[69,72,73]). For instance, Fig. 6 shows a relevant demographic measure, the size of the largest particle cluster, as well as pictures of the LCS landscape at times of great demographic flux. Here, the combination of

LCSs and a fast increase in the largest cluster size are related, as are a fast decrease and the splitting of LCSs (Fig. 6a).

We can also use this logic to identify demographic shifts in reproducing populations that are almost certainly induced by the flow and not by biology. The probability that large, monotone demographic shifts arise due to biology can easily be calculated from the negative binomial distribution, assuming the best and worst possible scenario for replicators. Paying attention only to large swings which are highly improbable ($p < 0.005$) given the biology alone allows us to infer the effects of merging and division

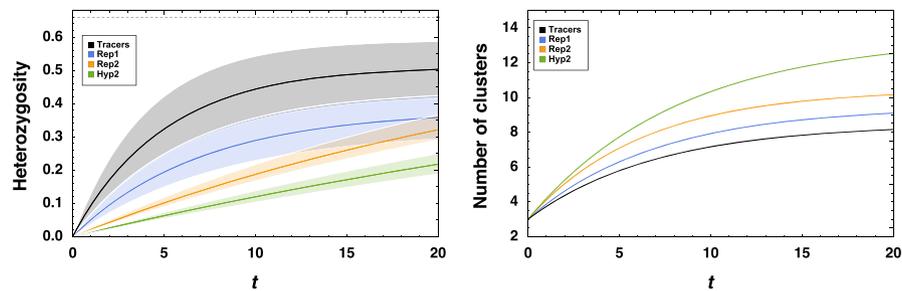

**Fig. 7 Dynamic population structure and small-scale chaotic migration effects increase cluster diversity.** Results of many simulations beginning with three dense clusters of particles near three distinct vortices of a seven-vortex flow, of which one realization is shown in Supplementary Fig. 1. The three clusters are considered to be "dyed" so that their lineages can be tracked over time. (Left-hand panel) The diversity of each cluster having at least 10 particles, as measured by the mean cluster heterozygosity (which here has the value 0 for a cluster of all one lineage and value 2/3 for a cluster with all three lineages represented equally). Different colors mark different metabolisms. The solid line represents the average over many simulations (see "Methods" section) and the opaque background represents one standard deviation. (Right-hand panel) The number of clusters consisting of 10 or more particles, averaged over all simulations, starting from the initial three vortices.

of structures without having to constantly calculate the Lyapunov exponent landscape.

Examining these extreme, flow-driven population shifts over many simulations generates the statistics shown in Fig. 6. While passive tracers provide a reasonable first approximation, especially for the frequency of division and merging events, the impact of the demographic shifts (as measured by the percentage change in the largest cluster) are decreased for more complicated metabolisms. This is likely due to the natural behavior of these systems to locally increase concentration gradients. It is clear from previous work[74,75] that analysis of the gradients in particle concentrations alone is inadequate for understanding systems undergoing a birth–death process, since true birth–death processes with motion (or, "superprocesses"[64,76–78]) tend to be much patchier and support larger, less smooth gradients in particle number. Therefore, as an initially spread-out population of cooperators coarsens into dense clusters, its ability to sense the boundaries of a coherent structure decreases; as a consequence, it is not always affected if a section of a coherent structure it is not occupying is cleaved off.

**Chaotic flows induce small-scale migration events that begin new colonies.** In addition to the division and merging of well-defined LCSs explored above, particles which are trapped with one LCS can be captured by another, a process which we call "migration" in order to maintain the distinction with division. Migration events occur due to the imperfect trapping by permanent vortices and the very slight stochasticity arising from a particle's offspring being placed very close to, but not completely atop them; at times, regions of positive Lyapunov exponent can move into an otherwise perfectly trapping LCS without splitting it, but causing some particles to bleed out onto the unstable manifold of the flow. These particles can then sometimes end up in a different, perhaps unpopulated LCS. Put another way, migration refers to the changes in population structure that occur due to the chaotic, rather than structured, part of the flow. Examples of migratory events can be seen in Supplementary Movie 3, which show examples of the successful replication and spread of small inocula of Replicase R2.

The distinction between migration and division is important, because whereas in the previous section, where division of one non-permanent coherent structure into two or more leads to a division of roughly equal fractions of particles, migration events do not involve the creation or annihilation of LCSs and involve only small changes in the demographics of the parent LCS. However, these small fractions were often seen to successfully seed an empty LCS, growing to much larger fractions and thereby promoting their lineage disproportionately, somewhat akin to "gene surfing" seen in other studies[79].

For instance, Supplementary Fig. 1 shows a realization in which a seven-vortex system initially seeded with a large inoculum of a three-species hypercycle dyed by lineage spreads via splitting, merging, and migration. By focusing only on the population very close to the seven point vortices (the neighborhood around point vortices almost always includes very small LCSs), we ensure that diversity spreading via splitting is not tracked. Migration events eventually lead to the satellite LCSs of all seven vortices being populated, and also contribute to the spread of different lineages, contributing to genetic diversity. Figure 7 illustrates the average number of clusters (interacting components, equivalent to components of the geometric graph[80] induced on the population by $R_{int}$) in different metabolisms over time, as well as the gradual mixing of different lineages, as measured by the mean cluster heterozygosity ($H = 1 - \sum_{l=1}^{3} f_l^2$, where $f_l$ represents the frequency of lineage $l$ in a cluster).

While the inherent patchiness of replicating metabolisms (as discussed in previous sections) leads to a higher number of particle clusters in a rather trivial way, the evolution of heterozygosity in living populations when compared to passive tracers is quite surprising. Passive tracers exhibited by far the greatest heterozygosity, and in some simulations exhibited a nearly perfect mixing of lineages after a short time. Living populations, on the other hand, had much less diversity on average. However, this can also be understood in terms of the patchiness that arises from the "death-anywhere, birth-locally" property of our process. This introduces number fluctuations at two levels not present in passive tracers. Firstly, the number of lineages is not guaranteed to be preserved, as stochastic effects can even eliminate a whole lineage, drastically decreasing the maximum possible heterzygosity. Additionally, the "gene surfing" effect mentioned above guarantees that newly formed clusters have zero heterozygosity, and demographic fluctuations even in a large cluster would tend to diminish heterozygosity. The balance between the role of the fluid and of the biology in determining genetic variation is therefore a rich and interesting one.

## Discussion

In this work, we have focused on qualitative features of turbulent flows and their similarity to theoretical notions of compartments in biological modeling. In such a generic setting, it is impossible to try to achieve exact quantitative results, as the types of biological metabolisms, cooperative mechanisms, and flow-kinetic parameters could vary widely. Furthermore, the role of additional physical effects in fluid environments, such as stratification, would depend dramatically on the particular environment of interest.

While the parameter space for our model is indeed quite large, we suspect that reasonable variations in many parameters (such as the difference between the death rate and birth rate of particles) will not cause qualitative differences in our results. However, there is one critical parameter, the Damköhler number $Da$, which controls the evolution of populations. Our work suggests that, similar to what has been seen in other works combining flow/migration and biological reproduction[62,65], an $\mathcal{O}(1)$ Damköhler number is optimal for catalytic cooperation. Outside this regime, the effect of the flow seems to be highly pessimal for cooperative metabolisms, suggesting that physical compartmentalization would be required.

However, within this regime, the flow itself can mimic important properties of compartments. These are not limited to kinematically isolating cooperators from parasites. Coherent structures in this regime can also provide many of the benefits associated with membranes, including merging and splitting of colonies, facilitating the spread and combination of genetic material. We therefore believe (in)compatibility of fluid and biological timescales in different aqueous environments could have acted as a strong selective force not only on the chemistry of the early metabolisms themselves, but also on the need for and development of physical compartments.

We believe that this work will establish a fundamentally new direction in the study of population structure in aqueous environments. Specific conditions relevant to origin of life problems, e.g. surface vents ("warm little ponds"), fast streams, and oceanic currents, can be studied in further detail (and with better estimates of the Damköhler number) to investigate the suitability of these environments for cooperative molecules and cells. In each of these environments, the inclusion of a third spatial dimension can complicate, but also enrich the models presented here. Upwellings, which we did not consider here, could for instance can benefit diversity by introducing new species and new genetic material. We hope that our work can inspire more scrutiny into the physical role of water in enabling cooperation, both by theorists and experimentalists and thereby enable more simplistic models of origin of life.

## Methods

**Simulations**. Simulations were performed in MATLAB by combining flow kinematics, incorporated via the function ode23, and a stochastic birth–death process implemented via a time-dependent Gillespie simulation[81]. In more detail, the simulations update the population according to the following steps:

- The system is initialized with a certain number of each particle species localized in the unit torus at positions picked uniformly-at-random (in Fig. 4, with distance from the vortex core not exceeding 1/50). Units of time are measured in a frame for which the death rate $d$ for particles is $d = 1$.
- The positions of all vortices and particles are updated via the function ode23 until either a stochastic event occurs (see below) or 0.01 units of time has elapsed since the beginning of the fluid motion. Stopping every 0.01 units of time in the absence of a stochastic reaction ensures that errors in the propensity vector (see below) do not accumulate by mis-apprehending which particles should be considered catalyzed.
- Between fluid motions, the distance between particles is assessed via the MATLAB function knnsearch. Depending on the metabolism used, particles which receive a catalytic boost from a neighbor are assigned one accordingly.
- Stochastic birth–death events operate according to a Gillespie process, adjusted for the fact that the rates (namely, the birth rate, due to time-dependent catalyzation) are time-dependent. Between fluid motions, when the state of catalysis for particles is assessed, a local propensity vector $p = [\sum_i^n s_i + \beta \iota_i, nd]$ is generated, where $n$ is the number of existing particles, $s_i$ is the self-replication rate, and $\iota_i$ is the indicator function on whether particle $i$ has received cooperative support. For all simulations, the values $d = 1$, $\beta = 0.7$, $s_i = 1.5$ (for replicase particles in the R1 replicase), $s_i = 0.8$ (all other particles) were used. The propensity, which consists of the sum rates for birth and death (the only stochastic reactions) is updated whenever an event occurs or the motion stops. According to standard techniques[81], the sum of propensities is normalized and integrated in time until the appropriate hitting time $\tau \sim \text{Exp}(1)$ for an event is reached. An event is chosen according to the probability distribution of births and deaths at the hitting time—therefore, each event consists of only one death or one reproduction, with the death of a particular individual occurring with probability $d/(nd + \sum_i^n s_i)$ and reproduction of a particular individual occurring with probability $s_i/(nd + \sum_i^n s_i)$. If an individual is chosen to reproduce, their offspring takes their species type and is placed uniformly-at-random no further than 1/200 away from their parent.

**Collocalization**. For this initial section, exploring only the role of a single LCS in facilitating the important properties of "coming together" and "staying together", we used an unsteady double-gyre, a flow commonly used as a benchmark for vortex recognition and FTLE calculation algorithms:

$$\psi(x, y, t) = A \sin(\pi f(x,t)) \sin(\pi y) \quad (4)$$

$$f(x,t) = \varepsilon \sin(\omega t) x^2 + (1 - 2\varepsilon \sin(\omega t)) x \quad (5)$$

where $A = 0.5$ and $\omega = 2\pi$ were used and $\varepsilon$ ranged from $10^{-2}$ to $10^0$. Although this system consists of two vortices (so that the total circulation in the system is zero), we are mainly interested in the confining force of one LCS as the amount of noise is increased, and so we simply measured the residence time as the time in which a passive tracers stayed in one half of the system if initialized in the center of one LCS.

For all metabolisms, 10,000 simulations were performed for differing values of $\varepsilon$ and $R_{\text{int}}$, the radius of possible cooperative interaction. For each metabolism, we initialized an inoculum of five particles per species type at positions chosen uniformly-at-random near (within a radial distance of 1/20) the vortex core ($r = 0$). If the population increased by a factor of 1000, the simulation was considered a success; if the population died out completely, the simulation was considered a failure. The sum total of successes, divided by 10,000, yields the numbers plotted in Fig. 4.

**Point-vortex flow**. Because our simulations already involve some operations which can be quite costly as the population grows (for instance, knnsearch takes $\mathcal{O}(n \log n)$ time for a population of size $n$), we wanted to employ a model of turbulence which could generate an infinite number of distinct (yet statistically identical) flows in an efficient manner. We did not concern ourselves with more complicated physics (such as influx of energy by forcing or outflux by viscous dissipation), instead opting for a perfectly conservative fluid with a perfectly conserved number of point vortices, whose motion is given by Eqs. (2) and (3), where $\Gamma_j$ is the circulation of the $j$th entity, having value of either $+1$ or $-1$ for a point vortex and 0 for a particle, and $\mathbf{d}_{ij}$ is the $\mathcal{L}_2$ distance between entity $i$ and $j$. The ramifications of these equations is that both particles and point vortices are simply advected by the linear superposition of the vorticity arising from all vortices in the domain.

In all sections we fixed a number of point vortices (6 or 7) on a doubly periodic domain, with half having positive handedness ($\Gamma = +1$) and half having negative handedness ($\Gamma = -1$). In the case of an odd number of vortices, a coin was flipped to determine the handedness of the final vortex. Because the domain is doubly periodic, determining the flow involves summing an infinite number of image vortices[82], such that Eqs. (2) and (3) become

$$\frac{d}{dt}x_i = \sum_{j=1}^{N} \Gamma_j \sum_{m=-\infty}^{m=\infty} \frac{-\sin(y_i - y_j)}{\cosh(x_i - x_j - 2\pi m) - \cos(y_i - y_j)},$$

$$\frac{d}{dt}y_i = \sum_{j=1}^{N} \Gamma_j \sum_{m=-\infty}^{m=\infty} \frac{\sin(x_i - x_j)}{\cosh(y_i - y_j - 2\pi m) - \cos(x_i - x_j)}.$$

Typically, one would have to employ a technique such as Ewald summation to deal with the doubly infinite sum over $m$. However, the reader can easily check that the size of the terms drops off incredibly fast, with the size of the $|m| = 2$ terms already being $\mathcal{O}(10^{-20})$ or smaller. We therefore truncated the sum at $|m| = 2$.

The number of vortices allows for rich dynamics, as a generic realization produces elliptic structures at (at least) three different scales: near a vortex, between two vortices (of any sign combination), and within short-term bound four-vortices. As seen in the main text, we often observed long-term elliptic structures that fell into none of these three categories as well.

**Identifying coherent structures**. The identification of LCSs in Figs. 4a, 5a, and 6a was performed using MATLAB code provided by the http://dabirilab.com/software/Dabiri Lab (LCS Matlab Kit v2.3)[68,83]. In calculating the FTLE, the velocity field on a 50 × 50 mesh was calculated at intervals of $\Delta t = 0.1$ from the positions of the vortices, and 15 such timesteps were integrated to calculate the FTLE plotted in the figure. Contour plots delimiting basins of attraction (non-positive FTLE) were then plotted in Mathematica.

We note that the code used was not designed to be employed on a doubly periodic surface, and therefore we believe that the values of the FTLE reported at the gridpoints closest to the domain boundary are incorrect. However, the interior LCSs indicated in Figs. 4a, 5a, and 6a were also identified by a different algorithm[69], also from the Dabiri group, which colors structures based on the kinematic similarity of passive tracers trajectories (generated for the same flow whose field was originally used to create Figs. 4a, 5a, and 6a).

**Escaping parasites**. To test whether LCSs would allow spatially dependent metabolisms to escape from parasites in a manner similar to proto-membranes, we simulated the Replicase R2 metabolism, which represents a cooperator and a parasite. To check if their relative fitness was affected by the flow, we examined a Wright–Fisher-type process instead of our usual branching process. The Wright–Fisher process employed has a fixed population size of $N = 200$, with the entire population updating at fixed discrete time intervals $T_{upd}$, at which time a new generation is formed by picking, for each of the 200 members, a parent with probability proportional to the parent's relative fitness in their population. The initial condition is distributed uniformly-at-random.

The Wright–Fisher simulations were continued until either all parasites were extinct, or the whole population (including parasites, which cannot survive once all healthy particles have been extinguished) were extinct. The former case was considered a success, the latter case a failure. The number of instances of the latter, divided by 10,000, gives the points plotted in Fig. 5.

We then reverted to our branching process, starting from a population of 50 per species localized near the core of one out of seven vortices, of which a fraction $f = 1/2$ were chosen uniformly-at-random to be parasites. Once the population grew to a size of 5000, we recorded the average fraction of cooperating replicases. Performing this over 1000 simulations gives the histogram in Fig. 5.

**Division and merging**. Calculating the FTLE field is quite computationally costly, and counting elliptic LCSs without fear of error would require an extremely high spatial resolution, and therefore making full evaluation of the flow properties in a simulation is a much more involved process than a full evaluation of the biological properties. Furthermore, it is easy to predict when demographic changes are not due to the biological process from this evaluation. For instance, we can assign a minimum and maximum probability to the number of birth–death events occurring within a certain timespan given the approximate population size, and can also predict the extremal probabilities of the outcome in which all (or most) such birth–death events were either birth or death. The latter is particularly easy; the likelihood of $M$ death events in a row before a birth must be bounded above by NB$(1, d/(s_i + d), M)$, where NB$(a,b,c)$ denotes the value of the negative binomial distribution giving the probability of $c$ deaths before 1 birth, given that the probability of a death in a population of size $n$ where no individual is catalyzed (hence, the situation in which the probability of death is always maximized, giving the maximal probability of $M$ deaths before a birth) is $dn/(ns_i + nd)$. Using our standard kinetic rates, a sequence of 10–15 deaths without a birth is already an extremely unlikely event, so that a dip in the largest cluster size of 10–15 would indicate a flow-based event (division of coherent structure) with high probability. In each of 1000 simulations, we paid attention only to monotonic changes in the size of the largest cluster with absolute value bigger than 10. The statistics of these changes are plotted in Fig. 6.

**Migration and diversity**. For Supplementary Fig. 1, seven vortices were used and were assigned an integer label which was maintained throughout the simulations. The curves shown were generated by simulating 1000 runs per metabolism, initializing 100 of each particle species in the vortices labeled 1–3 (therefore implicitly assuming that the metabolism had experienced some reproductive success in one vortex). Simply tracking the dynamics of particles from the three original lineages over time generated the left-hand side of Fig. 7. The right-hand side was given by counting the number of components (with 10 or more particles included) in a geometric graph[80] induced by drawing an edge between two particles iff they were at a distance (in the Euclidean metric) $< R_{int}$.

**Reporting summary**. Further information on research design is available in the Nature Research Reporting Summary linked to this article.

## Data availability
No additional data was used besides the results of numerical simulations using the parameters described in the text. Additional summary statistics of the data plotted may be available upon reasonable request.

## Code availability
MATLAB scripts to reproduce the results in the manuscript can be found on GitHub: https://github.com/ski-krieger/EarlyLife.

## Acknowledgements
The authors thank Jeff Gerold, Kamran Kaveh, Carl Veller, Alison Hill, Abigail Plummer, Greg Wagner, and Alex McAvoy for insightful contributions.

## Author contributions
M.S.K. designed the study and analytical models. S.S. and M.S.K. performed numerical simulations. M.S.K., S.S., and M.A.N. wrote the manuscript.

## Competing interests
The authors declare no competing interests.

## Additional information
**Supplementary information** is available for this paper at https://doi.org/10.1038/s41467-020-15780-1.

**Correspondence** and requests for materials should be addressed to M.S.K.

**Peer review information** *Nature Communications* thanks Norikazu Ichihashi and the other, anonymous, reviewers for their contribution to the peer review of this work.

**Reprints and permission information** is available at http://www.nature.com/reprints

**Publisher's note** Springer Nature remains neutral with regard to jurisdictional claims in published maps and institutional affiliations.